\documentclass[pra,twocolumn,showpacs]{revtex4}
\usepackage{graphicx}
\usepackage{amsmath}
\usepackage{color}

\begin{document}
\title{Combination of a magnetic Feshbach resonance \\ and an optical bound-to-bound transition}
\author{Dominik M. Bauer}
\author{Matthias Lettner}
\author{Christoph Vo}
\author{Gerhard Rempe}
\author{Stephan D\"{u}rr}
\affiliation{Max-Planck-Institut f{\"u}r Quantenoptik, Hans-Kopfermann-Stra{\ss}e 1, 85748 Garching, Germany}

\begin{abstract}
We use laser light near resonant with an optical bound-to-bound transition to shift the magnetic field at which a Feshbach resonance occurs. We operate in a regime of large detuning and large laser intensity. This reduces the light-induced atom-loss rate by one order of magnitude compared to our previous experiments [D.~M.\ Bauer {\it et al.} 
Nature Phys. {\bf 5}, 339 (2009)]. The experiments are performed in an optical lattice and include high-resolution spectroscopy of excited molecular states, reported here. In addition, we give a detailed account of a theoretical model that describes our experimental data.
\end{abstract}

%\date{\today}
\hyphenation{Feshbach}


% 34.50.Cx    Elastic; ultracold collisions
% 33.40.+f    Multiple resonances (including double and higher-order resonance processes, such as double nuclear magnetic resonance, electron double resonance, and microwave optical double resonance) (see also 76.70.-r Magnetic double resonances and cross effects in condensed matter)
% 33.20.Kf    Visible spectra 
% 03.75.Hh    Static properties of condensates; thermodynamical, statistical, and structural properties 

\maketitle

\section{Introduction}
Many properties of ultracold gases are determined by the interparticle interaction which is characterized by the $s$-wave scattering length $a$. This makes it desirable to tune this parameter. A much-used method for this purpose is a magnetic Feshbach resonance \cite{moerdijk:95,inouye:98,chin:0812.1496}. An alternative method is a photoassociation resonance, which is sometimes also called optical Feshbach resonance \cite{fedichev:96a,bohn:97,fatemi:00,theis:04,thalhammer:05,jones:06}. A major advantage of photoassociation resonances is that the light intensity can be varied on short length and time scales, thus offering more flexible experimental control over the scattering length. The problem with photoassociation resonances is that the light induces inelastic collisions between atoms which lead to rapid loss of atoms. This is why photoassociation resonances have only rarely been used to tune the scattering length \cite{theis:04,thalhammer:05}. A solution for this problem exists for alkali earth atoms where narrow intercombination lines allow for tuning of $a$ with only moderate loss \cite{ciurylo:05,enomoto:08}. But this is not feasible in the large number of experiments with alkali atoms.

\begin{figure}[t!]
\includegraphics{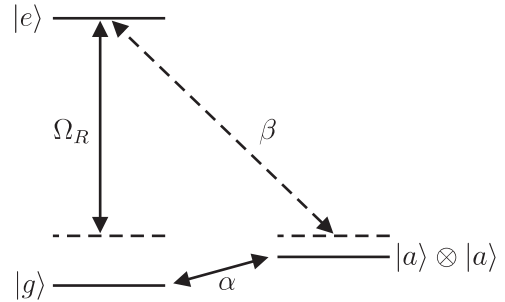}
\caption{
\label{fig-scheme}
Level scheme. Pairs of atoms, each in state $|a\rangle$, are coupled to a dimer state $|g\rangle$ belonging to the electronic ground state. This coupling has a strength $\alpha$ and causes a magnetic Feshbach resonance. Laser light is applied to drive a bound-to-bound transition from state $|g\rangle$ to an electronically excited dimer state $|e\rangle$ with Rabi frequency $\Omega_R$. This causes an ac-Stark shift (not shown) of state $|g\rangle$ which leads to a shift of the magnetic field at which the Feshbach resonance occurs. Typically, the same light can also drive photoassociation from state $|a\rangle\otimes|a\rangle$ to state $|e\rangle$ with coupling strength $\beta$. The photoassociation is a nuisance for the scheme used here. The light frequency is typically somewhat detuned from both transitions.
}
\end{figure}

In a recent experiment \cite{bauer:09} we explored an alternative scheme for controlling the scattering length with laser light. This scheme uses the existing coupling between an atom-pair state $|a\rangle\otimes|a\rangle$ and a molecular state $|g\rangle$ near a Feshbach resonance, as illustrated in Fig.\ \ref{fig-scheme}. By adding a light field that is somewhat detuned from a bound-to-bound transition between state $|g\rangle$ and an electronically excited molecular state $|e\rangle$, one can induce an ac-Stark shift of state $|g\rangle$. This results in a shift of the magnetic field $B_{\rm res}$ at which the Feshbach resonance occurs. If the magnetic field $B$ is held close to the Feshbach resonance, then spatial or temporal variations of the light intensity affect the scattering length. This scheme also suffers from light-induced inelastic collisions. However, in Ref.\ \cite{bauer:09} we reported a two-body loss rate coefficient as small as $K_2\sim 10^{-11}$ cm$^3$/s at parameters where the real part of the scattering length is changed by ${\rm Re}(a)/a_{\rm bg}-1=\pm1$ with respect to its background value $a_{\rm bg}$. This represents a reduction of the loss rate by one order of magnitude compared to a photoassociation resonance in $^{87}$Rb where $K_2\sim 10^{-10}$ cm$^3$/s was reported for the same change of ${\rm Re}(a)$ \cite{theis:04,thalhammer:05}. 

Here we show experimentally that a value as low as $K_2\sim 10^{-12}$ cm$^3$/s at ${\rm Re}(a)/a_{\rm bg}-1=1$ can be reached with our scheme, thus reducing the loss rate by one more order of magnitude compared to our previous results. This improvement is achieved by increasing the detuning and the intensity of the laser light, which is a commonly used method for reducing incoherent rates in experiments, which rely on the ac-Stark shift. When following this approach, two new issues have to be addressed. First, the detunings are no longer small compared to the typical splitting between the various hyperfine and magnetic substates of the excited state. Achieving a large detuning with respect to all excited states thus requires knowledge of the positions of all nearby excited states, so that excited-state spectroscopy has to be performed, on which we report in Sec.\ \ref{sec-spectroscopy}. Second, in our previous experiment \cite{bauer:09} we used a model for the temporal evolution of the cloud size to determine ${\rm Re}(a)$ and $K_2$. If the laser that drives the bound-to-bound transition is now operated at much larger power, then it creates a noticeable dipole trap. Its rapid turn-on induces large-amplitude oscillations of the cloud size which are difficult to model. We therefore explore an alternative way of measuring ${\rm Re}(a)$ and $K_2$. To this end, we pin the positions of the atoms with a deep optical lattice which minimizes the effect of the additional dipole trap. In the lattice, we use excitation spectroscopy and a loss measurement to determine ${\rm Re}(a)$ and $K_2$, respectively. These results are reported in Sec.\ \ref{sec-shift}.

In Ref.\ \cite{bauer:09} we also studied the behavior of the system when the laser light is tuned close to the bound-to-bound resonance. In this regime, we observed an Autler-Townes doublet in the loss rate coefficient as a function of magnetic field $K_2(B)$. In the same work, we presented a systematic study of the dependence of the positions $B_{\rm res}$ of the loss resonances on laser power and detuning. In Sec.\ \ref{sec-exp-Autler} of the present paper, we complement these measurements with a systematic study of the height and width of the resonances in this Autler-Townes doublet. These experimental data agree well with the theoretical model that we use, thus showing that the relevant physics is well understood.

Before turning to the experiment, we begin in Sec. \ref{sec-theory} with a detailed discussion of the theoretical model that we use to describe our experiments. 

\section{Theory}
\label{sec-theory}

We consider a system with three internal states, as shown in Fig.\ \ref{fig-scheme}. State $|a\rangle$ represents a single unbound atom. States $|g\rangle$ and $|e\rangle$ represent a dimer in the electronic ground and excited state, respectively. The atom-molecule coupling between the atom-pair state $|a\rangle\otimes|a\rangle$ and the dimer state $|g\rangle$ relevant for the Feshbach resonance is described by a coupling strength $\alpha$. Laser light couples states $|g\rangle$ and $|e\rangle$ on a bound-to-bound transition with a Rabi frequency $\Omega_R$. The same laser light drives photoassociation from state $|a\rangle\otimes|a\rangle$ to $|e\rangle$ characterized by a coupling strength $\beta$.

This model is closely related to previous studies of the combination of a Feshbach resonance with a photoassociation resonance, see {\it e.g.}\ Refs.\ \cite{abeelen:98,junker:08,mackie:08}. Unlike those references, we are mostly interested in the Feshbach resonance and the bound-to-bound transition. The photoassociation is a nuisance in our scheme, because any useful change of ${\rm Re}(a)$ that it induces is inevitably accompanied by the loss rates that limited the photoassociation experiments in Refs.\ \cite{theis:04,thalhammer:05}. Luckily, we find excited states in our experiment for which photoassociation is negligible.

Before turning to a quantitative model, we present a qualitative argument that motivates why creating a given change in ${\rm Re}(a)$ with our scheme causes a smaller loss rate than a photoassociation resonance would do. To explain this, we must first understand the limitation of the photoassociation resonance. If an infinite amount of laser intensity were available, then the value of $K_2$ caused by a photoassociation resonance could be reduced without modifying the change in ${\rm Re}(a)$. To this end, one would simply have to increase $\Delta_e$ along with the laser intensity, see Eq.\ (\ref{PA}). The same improvement could be achieved if a photoassociation resonance with a larger transition matrix element was available because that would be equivalent to having more laser intensity.

Our scheme relies on the fact that the typical interatomic distance is orders of magnitude larger in the atomic gas than within a single molecule. As a result, the transition matrix element is typically orders of magnitude smaller for a photoassociation resonance than for a bound-to-bound transition, see Eq.\ (\ref{beta-small}). Hence at a given laser intensity, one can detune the laser pretty far in our scheme and still achieve a significant light-induced change in ${\rm Re}(a)$, whereas a photoassociation resonance driven with the same laser intensity requires a smaller laser detuning, which results in a larger loss rate.

\subsection{Hamiltonian}

According to Refs.\ \cite{kokkelmans:02,koehler:03} the system with the light off is described by the Hamiltonian
\begin{eqnarray}
\hat H 
&=& \sum_{j\in\{a,g,e\}} \int d^3{\bf x} \hat \Psi_j^\dag ({\bf x}) H_j ({\bf x}) \hat \Psi_j ({\bf x}) 
\nonumber \\ &+& 
\frac12 \int d^3{\bf x}_1 d^3{\bf x}_2 \hat \Psi_p^\dag ({\bf x}_1,{\bf x}_2) U_{{\rm bg}}({\bf x}_{12})
\hat \Psi_p ({\bf x}_1,{\bf x}_2) 
\nonumber \\ &+& 
\hbar \int d^3{\bf x}_1 d^3{\bf x}_2 \left[ \hat \Psi_g^\dag ({\bf X}_{12}) \alpha({\bf x}_{12})
\hat \Psi_p ({\bf x}_1,{\bf x}_2) + {\rm H.c.} \right]
\nonumber \\
\end{eqnarray}
with the bosonic field operators $\hat \Psi_j({\bf x})$ for $j\in\{a,g,e\}$, the single-particle Hamiltonians $H_j({\bf x})$, the relative coordinate ${\bf x}_{12}={\bf x}_1-{\bf x}_2$, the center-of-mass coordinate ${\bf X}_{12}=({\bf x}_1+{\bf x}_2)/2$, and the abbreviation $\hat \Psi_p({\bf x}_1,{\bf x}_2)=\hat \Psi_a({\bf x}_1) \hat \Psi_a({\bf x}_2)$ for the annihilation of an atom pair. Elastic two-atom collisions far away from the Feshbach resonance are described by $U_{\rm bg}({\bf x}_{12})$ and the atom-molecule coupling relevant for the Feshbach resonance is described by $\alpha({\bf x}_{12})$.

Collisions between more than two atoms were neglected here assuming that the interatomic potentials are short ranged compared to the average interatomic distance. Furthermore, collisions involving at least one molecule were neglected, assuming that the density of molecules is low.

$\alpha({\bf x}_{12})$ is the Fourier transform of \cite{timmermans:99}
\begin{eqnarray}
\label{def-alpha}
\alpha({\bf k}) = \frac1 \hbar \sqrt{\frac V2} \langle g|H| {\bf k} \rangle
,
\end{eqnarray}
where $V$ is the quantization volume, $|\bf k\rangle$ is a two-atom scattering state that has an incoming plane wave with wave vector $\bf k$ in the relative coordinate, and $H$ is the Hamiltonian describing the atom-molecule coupling in the collision of two atoms. Its position representation has the asymptotic form
\begin{eqnarray}
\langle {\bf x}_{12}|{\bf k}\rangle
= \frac{e^{i{\bf kx}_{12}} + f({\bf k}) e^{ikx_{12}}/x_{12}}{\sqrt V}
, \quad x_{12}\to \infty
,
\end{eqnarray}
where $f({\bf k})$ is the scattering amplitude. Combination of the last two equations shows that $\alpha({\bf k})$ is independent of $V$.

If all incoming particles in the gas are slow enough, then we can replace $\alpha({\bf k})$ by a constant $\alpha=\lim_{{\bf k}\to0}\alpha({\bf k})$. In the position representation, this corresponds to a contact potential in the relative coordinate. We use an analogous approximation for $U_{\rm bg}$ and obtain $\hat H=\int d^3 {\bf x} {\cal H}({\bf x})$ with the Hamiltonian density \cite{timmermans:99}
\begin{eqnarray}
\label{cal-H}
{\cal H}({\bf x})
&=& \sum_{j\in\{a,g,e\}} \hat \Psi_j^\dag ({\bf x}) H_j ({\bf x}) \hat \Psi_j ({\bf x}) 
+ \frac{U_{{\rm bg}}}2 \hat \Psi_p^\dag ({\bf x}) \hat \Psi_p ({\bf x}) 
\nonumber \\ &+& 
\hbar \left[ \hat \Psi_g^\dag ({\bf x}) \alpha \hat \Psi_p ({\bf x}) + {\rm H.c.} \right]
.
\end{eqnarray}
Note that $U_{\rm bg} = 4\pi\hbar^2 a_{\rm bg}/m$ is related to the background scattering length $a_{\rm bg}$ and the atomic mass $m$.

We now extend this model to include effects caused by the light, which has an electric field of the form $E=-E_0 \cos(\omega_L t)$ with amplitude $E_0$ and angular frequency $\omega_L$. In analogy to Eq.\ (\ref{cal-H}) we obtain the following additional terms for the Hamiltonian density
\begin{eqnarray}
{\cal H}_L({\bf x})
&=& \hbar \cos(\omega_L t) \left[ \hat \Psi_e^\dag ({\bf x}) \Omega_R \Psi_g ({\bf x}) + {\rm H.c.} \right]
\nonumber \\ &+&
2 \hbar \cos(\omega_L t) \left[ \hat \Psi_e^\dag ({\bf x}) \beta \hat \Psi_p ({\bf x}) + {\rm H.c.} \right]
.
\end{eqnarray}
The Rabi frequency $\Omega_R$ and the coupling strength $\beta$ describe the bound-to-bound transition and the photoassociation, respectively. $\Omega_R$ is related to a matrix element of the matter-light interaction term in the Hamiltonian, which in the electric dipole approximation reads $H(t)=-dE(t)$. This yields $\hbar \Omega_R \cos(\omega_L t) = - \langle e|d|g\rangle E(t)$. We abbreviate $d_{eg}=\langle e|d|g\rangle$ and obtain the well-known relation 
\begin{eqnarray}
\Omega_R = \frac{E_0 d_{eg}}\hbar
.
\end{eqnarray}
For $\beta$ we obtain in analogy to Eq.\ (\ref{def-alpha})
\begin{eqnarray}
\beta = \frac{E_0}{2\hbar} \sqrt{\frac V2} \lim_{k\to0} \langle e|d|{\bf k}\rangle
.
\end{eqnarray}

Finally, we specify the single-particle Hamiltonians which consist of a kinetic part and an internal part $E^{\rm int}_j$
\begin{eqnarray}
H_j({\bf x}) = - \frac{\hbar^2\nabla^2}{2m_j} + E^{\rm int}_j
\end{eqnarray}
for $j\in\{a,g,e\}$. Here $m_j=m(2-\delta_{aj})$ is the mass of a particle in state $j$ and $\delta_{ij}$ is the Kronecker symbol. The internal energy of each state depends nonlinearly on the magnetic field $B$. Near the pole of the unshifted Feshbach resonance $B_{\rm pole}$ we approximate this dependence as linear and obtain
\begin{eqnarray}
E^{\rm int}_j = - \mu_j (B-B_{\rm pole}) + \hbar \omega_{eg} \delta_{ej}
,
\end{eqnarray}
where $\mu_j$ is a magnetic dipole moment. At $B=B_{\rm pole}$, the internal states $|a\rangle$ and $|g\rangle$ are degenerate whereas the internal state $|e\rangle$ has an energy offset $\hbar \omega_{eg}$. 

\subsection{Mean-Field Model}
We assume that the population in each state is Bose condensed and can be described by a mean field $\psi_j({\bf x})=\langle\hat\Psi_j({\bf x})\rangle$. In addition, we assume that the system is homogeneous. We take the expectation value of the Heisenberg equation of motion $i\hbar d\hat\Psi_j/dt=[\hat\Psi_j,H]$ and approximate the field operators as uncorrelated, so that the expectation values factorize \cite{timmermans:99}. Physics beyond this approximation is discussed in Refs.\ \cite{holland:01a,koehler:03}. We obtain
\begin{subequations}
\begin{eqnarray}
i \frac{d}{dt} \psi_a &=& \frac{E^{\rm int}_a}\hbar \psi_a + \frac{U_{\rm bg}}\hbar |\psi_a|^2 \psi_a 
\nonumber \\ && 
+ 2 \alpha^* \psi_a^* \psi_g + 4 \beta^* \psi_a^* \psi_e \cos(\omega_L t)
\\
i \frac{d}{dt} \psi_g &=& \alpha \psi_a^2 + \frac{E^{\rm int}_g}\hbar \psi_g + \Omega_R^* \psi_e \cos(\omega_L t) \\
i \frac{d}{dt} \psi_e &=& (2 \beta \psi_a^2 + \Omega_R \psi_g) \cos(\omega_L t)
+ \frac{E^{\rm int}_e}\hbar \psi_e 
.
\end{eqnarray}
\end{subequations}
We move to an interaction picture by replacing $\psi_e \to \psi_e e^{i\omega_Lt}$ and perform a rotating-wave approximation by neglecting coefficients rotating as $e^{\pm2i\omega_Lt}$. We then move to another interaction picture by replacing $\psi_j\to \psi_j \exp((2-\delta_{aj})iE_a^{\rm int}t/\hbar)$ for $j\in\{a,g,e\}$. Hence (see also Ref.\ \cite{bauer:09})
\begin{subequations}
\label{mean-field}
\begin{eqnarray}
\label{mean-field-1}
i \frac{d}{dt} \psi_a &=& \frac{U_{\rm bg}}\hbar |\psi_a|^2 \psi_a 
+ 2 \alpha^* \psi_a^* \psi_g + 2 \beta^* \psi_a^* \psi_e
\\
\label{mean-field-2}
i \frac{d}{dt} \psi_g &=& \alpha \psi_a^2 + \Delta_g \psi_g + \frac12 \Omega_R^* \psi_e
\\
\label{mean-field-3}
i \frac{d}{dt} \psi_e &=& \beta \psi_a^2 + \frac12 \Omega_R \psi_g + \left( \Delta_e -\frac i2 \gamma_e \right) \psi_e 
,
\end{eqnarray}
\end{subequations}
where we abbreviated
\begin{subequations}
\begin{eqnarray}
\label{delta-g}
\Delta_g &=& \frac1\hbar \mu_{ag} (B-B_{\rm pole}) \\
\label{delta-e}
\Delta_e &=& - \Delta_L + \frac1\hbar \mu_{ae} (B-B_{\rm pole}) \\
\Delta_L &=& \omega_L-\omega_{eg}
\end{eqnarray}
\end{subequations}
with $\mu_{ag}=2\mu_a-\mu_g$ and $\mu_{ae}=2\mu_a-\mu_e$. In Eq.\ (\ref{mean-field-3}) we included an {\it ad hoc} decay rate $\gamma_e$ that represents spontaneous radiative decay from state $|e\rangle$ into states that are not included in the model, similar to Ref.\ \cite{abeelen:99}. A model similar to Eqs.\ (\ref{mean-field}) was used in Ref.\ \cite{mackie:08} to explain enhanced photoassociation loss rates near a magnetic Feshbach resonance \cite{junker:08}.

Note that the typical interatomic distance is orders of magnitude larger in the atomic gas than within a single molecule. For a typical excited state, this results in
\begin{eqnarray}
\label{beta-small}
|\beta\psi_a| \ll |\Omega_R|
.
\end{eqnarray}

\subsection{Adiabatic Elimination and Scattering Length}
\label{sec-a}
We assume that all the population is initially prepared in state $|a\rangle$ and that the populations in states $|g\rangle$ and $|e\rangle$ will remain small at all times so that they can be eliminated adiabatically, similar to Refs.\ \cite{abeelen:99,mackie:08}. This is a good approximation, e.g., if the angular frequencies $\alpha \psi_a$ and $\beta \psi_a$ are both small compared to $\Omega_R$ and $\gamma_e$ or compared to $\Delta_e$ and $\Delta_g$. This condition is always satisfied in the low-density limit, but for a very broad Feshbach resonance it might be difficult to reach this regime experimentally.

The adiabatic elimination is achieved by formally setting $(d/dt)\psi_g=(d/dt)\psi_e=0$. This is used to eliminate $\psi_g$ and $\psi_e$ from the equations. We obtain 
\begin{eqnarray}
i \frac{d}{dt} \psi_a &=& \frac{4\pi\hbar a}{m} |\psi_a|^2 \psi_a
\end{eqnarray}
with the complex-valued scattering length
\begin{eqnarray}
\label{a}
a &=& a_{\rm bg} - \frac{m}{2\pi\hbar} 
\nonumber \\
&\times& \frac{ |\alpha|^2 (\Delta_e - i \gamma_e/2) - {\rm Re}(\alpha^* \Omega_R^*\beta) + |\beta|^2 \Delta_g}{(\Delta_e - i \gamma_e/2) \Delta_g -|\Omega_R/2|^2}
.
\end{eqnarray}
The term ${\rm Re}(\alpha^* \Omega_R^* \beta)$ represents interference between the two possible ways to go from state $|a\rangle$ to state $|e\rangle$, either directly or indirectly through state $|g\rangle$. 

The real part of the scattering length is responsible for the mean-field energy \cite{duerr:09}. We assume that $a_{\rm bg}$ is real and obtain
\begin{eqnarray}
\label{Re-a}
&& \!\!\!\! {\rm Re}(a) = a_{\rm bg} - \frac{m}{2\pi\hbar} \ \frac1{(\Delta_g\Delta_e- |\Omega_R/2|^2)^2 +( \Delta_g\gamma_e/2)^2}
\nonumber \\ && \!\!\!\!
\times \Big[\Big(|\alpha|^2\Delta_e  \! - \! {\rm Re}(\alpha^* \Omega_R^*\beta) + |\beta|^2 \Delta_g\Big) 
\Big(\Delta_e\Delta_g-|\Omega_R/2|^2\Big)
\nonumber \\ && \!\!\!\!
+ |\alpha|^2\Delta_g(\gamma_e/2)^2 \Big]
.
\end{eqnarray}

The imaginary part of the scattering length gives rise to two-body loss with a rate equation \cite{duerr:09}
\begin{subequations}
\begin{eqnarray}
\frac{d n}{dt} &=& -K_2 n^2 g^{(2)} \\
K_2 &=& - \frac{8\pi\hbar}{m} {\rm Im}(a) 
\label{K2-def}
,
\end{eqnarray}
\end{subequations}
where $n=|\psi_a|^2$ is the atomic density, $K_2$ is the two-body loss coefficient for a Bose-Einstein condensate (BEC), and $g^{(2)}$ is the pair correlation function at zero relative distance. For a BEC with $N$ atoms $g^{(2)}=1-1/N$. Insertion of Eq.\ (\ref{a}) yields
\begin{eqnarray}
\label{K2}
K_2 = 2\gamma_e \frac{|\alpha \Omega_R/2|^2 - \Delta_g {\rm Re}(\alpha^* \Omega_R^*\beta) + |\beta|^2 \Delta_g^2}
{(\Delta_g\Delta_e- |\Omega_R/2|^2)^2 +( \Delta_g\gamma_e/2)^2}
.
\end{eqnarray}
All terms in the numerator are $\propto E_0^2$, so that the relative importance of the terms is independent of laser intensity. If $B$ is held near the unshifted Feshbach resonance, then $\Delta_g$ is small and all terms containing $\beta$ in Eq.\ (\ref{K2}) become negligible. Hence if one considers measurements of $K_2$ performed fairly close to the Feshbach resonance, then photoassociation is negligible and it is impossible to extract the value of $\beta$ only from such measurements (unless $\Omega_R$ vanishes).

The interference term in Eq.\ (\ref{a}) leads to a corresponding interference term in Eq.\ (\ref{K2}). Note that the minimum of $K_2$ observed in Ref.\ \cite{junker:08} is a result of destructive interference due to this term \cite{mackie:08}.

\begin{figure}[t!]
\includegraphics[width=0.9\columnwidth]{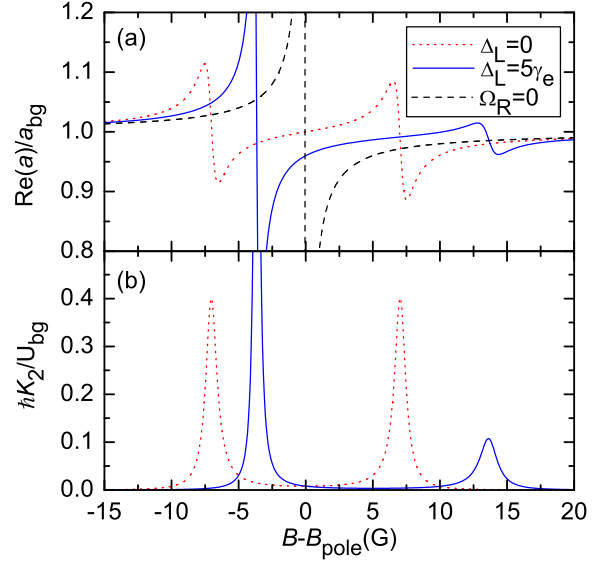}
\caption{
\label{fig-doublet}
(Color online) Predictions for ${\rm Re}(a)/a_{bg}$ and $K_2$ as a function of $B$ from Eqs.\ (\ref{Re-a}) and (\ref{K2}). For all curves, we choose $\hbar \gamma_e/\mu_{ae}=2$ G, $\beta=0$, and $\Delta B=0.2$ G; with $\Delta B$ defined in Eq.\ (\ref{alpha}). The dotted lines (red) are for resonant laser light $\Delta_L=0$ and $\hbar|\Omega_R|^2/\gamma_e\mu_{ag}=100$ G. They show two resonances that are symmetrically split around $B_{\rm pole}$. These resonances represent an Autler-Townes doublet. The solid lines (blue) are for large laser detuning $\Delta_L/\gamma_e=5$ and $\hbar|\Omega_R|^2/\gamma_e\mu_{ag}=100$ G. They also show two resonances, but their heights, widths, and distances from $B_{\rm pole}$ are quite different. The dashed line (black) is a reference without any light $\Omega_R=0$.
}
\end{figure}

Figure \ref{fig-doublet} shows predictions for ${\rm Re}(a)/a_{bg}$ and $K_2$ as a function of $B$. For large $|\Omega_R|$, one can clearly see two resonances in $K_2$ each of which is approximately Lorentzian. Each of these resonances is accompanied by a dispersive feature in ${\rm Re}(a)/a_{bg}$.

We show now that our model reproduces known results from the literature in the special cases of a pure Feshbach resonance or a pure photoassociation resonance. For a pure Feshbach resonance ($\beta=\Omega_R=0$) Eq.\ (\ref{a}) yields the familiar result \cite{moerdijk:95}
\begin{eqnarray}
\label{a-FR}
a = a_{\rm bg} \left( 1 - \frac{\Delta B}{B-B_{\rm pole}} \right)
\end{eqnarray}
with the width of the Feshbach resonance
\begin{eqnarray}
\label{alpha}
\Delta B = \frac{2 \hbar^2 |\alpha|^2}{U_{\rm bg} \mu_{ag}}
.
\end{eqnarray}
For a pure photoassociation resonance ($\alpha=\Omega_R=0$) Eq.\ (\ref{a}) yields
\begin{eqnarray}
a = a_{\rm bg} - \frac{m}{2\pi\hbar} \frac{|\beta|^2}{\Delta_e -i \gamma_e/2}
,
\end{eqnarray}
which is a Breit-Wigner form \cite{breit:36} as a function of $\Delta_L$ or $B$. The real and imaginary parts are
\begin{subequations}
\label{PA}
\begin{eqnarray}
{\rm Re}(a) &=& a_{\rm bg} -\frac{m}{2\pi\hbar} \Delta_e  \frac{|\beta|^2}{\Delta_e^2 +(\gamma_e/2)^2} \\
K_2 &=& 2\gamma_e \frac{|\beta|^2}{\Delta_e^2 +(\gamma_e/2)^2}
,
\end{eqnarray}
\end{subequations}
which is identical to Eq.\ (10) in Ref.\ \cite{bohn:97} in the limit $\Gamma_{\rm stim}\ll \Gamma_{\rm spon}$.

A quantitative comparison of Eq.\ (\ref{K2}) with our experimental data (here and in Ref.\ \cite{bauer:09}) shows that $\beta$ is negligible for most of the excited states $|e\rangle$ that we use. We therefore set $\beta=0$ for the rest of the theory section.

\subsection{Large Detuning}
\label{sec-large-detuning}

A good part of our experiments is performed in the limit of large laser detuning where $|\Delta_L|\gg |\mu_{ae} (B-B_{\rm pole})/\hbar|$ and $|\Delta_L|\gg\gamma_e$ and with $\beta=0$. In Eq.\ (\ref{a}) we can approximate $1/(\Delta_e-i\gamma_e/2)\sim -1/\Delta_L-i\gamma_e/2\Delta_L^2$ and obtain a Breit-Wigner form
\begin{eqnarray}
\label{a-Breit-Wigner}
a = a_{\rm bg} \left( 1- \frac{\Delta B}{B-B_{\rm res} - i W/2 } \right)
\end{eqnarray}
with $\Delta B$ from Eq.\ (\ref{alpha}). The real and imaginary parts are a dispersive line shape and a Lorentzian, respectively,
\begin{subequations}
\label{Re-a-Im-a}
\begin{eqnarray}
{\rm Re}(a) &=& a_{\rm bg}\left( 1- \frac{\Delta B(B-B_{\rm res})}{(B-B_{\rm res})^2 + W^2/4} \right)
\\
\label{K2-Lorentz}
K_2 &=& \frac{K_2^{\rm max}}{1+4(B-B_{\rm res})^2/W^2}
.
\end{eqnarray}
\end{subequations}
The resonance position $B_{\rm res}$, the maximum loss rate coefficient $K_2^{\rm max}$, and the full width at half maximum $W$ of the Lorentzian are given by
\begin{subequations}
\label{large-detuning}
\begin{eqnarray}
\label{large-detuning-K2-max}
K_2^{\rm max} &=& \frac{\hbar}{\mu_{ag}} \ \frac{8 |\alpha|^2}{W}
\\
\label{large-detuning-W}
W &=& \frac{\hbar}{\mu_{ag}} \ \frac{|\Omega_R|^2}{4\Delta_L^2} \gamma_e 
\\
\label{large-detuning-Bres}
B_{\rm res} - B_{\rm pole} &=& - \frac{\hbar}{\mu_{ag}} \ \frac{|\Omega_R|^2}{4\Delta_L}
.
\end{eqnarray}
\end{subequations}
The far-detuned bound-to-bound coupling yields the well-known ac-Stark shift of state $|g\rangle$ and this shifts $B_{\rm res}$.

As in most applications of ac-Stark shifts, we wish to achieve a certain value of $|\Omega_R|^2/\Delta_L$ and at the same time keep the rates for incoherent processes as low as possible. Hence, it is advantageous to increase the detuning and power of the laser in a way that keeps $|\Omega_R|^2/\Delta_L$ constant. This yields $W\to0$ and $K_2(B)\to (4\pi\hbar|\alpha|^2 /\mu_{ag}) \delta(B-B_{\rm res})$, where $\delta$ denotes the Dirac delta function. For any given value of $B\neq B_{\rm res}$ one can thus decrease $K_2(B)$ by increasing the detuning and the laser power sufficiently far.

In general, it is possible that several excited states contribute noticeably to $a$. Our model is easily adapted to this situation by introducing a separate version of Eq.\ (\ref{mean-field-3}) for each excited state and by including sums over the excited states in Eqs.\ (\ref{mean-field-1}) and (\ref{mean-field-2}). In the limit of large laser detuning and with $\beta=0$ for each excited state, Eqs.\ (\ref{a-Breit-Wigner})--(\ref{large-detuning-K2-max}) remain unchanged and a sum over the excited states appears on the right hand side of Eqs.\ (\ref{large-detuning-W}) and (\ref{large-detuning-Bres}).

\subsection{Autler-Townes Model for Weak Damping}
More insight into the physics of the problem can be gained from an Autler-Townes model \cite{autler:55,cohen-tannoudji:92}. In addition, analytic expressions for the position, height and width of the resonances in $K_2(B)$ can be derived.

This approach is based on the assumption that the dominant frequencies in the problem are
$\Omega_R$ and/or $(\Delta_g-\Delta_e)$. In this case, one can first diagonalize the driven two-level system spanned by $|g\rangle$ and $|e\rangle$ and subsequently treat the coupling to state $|a\rangle$ as a weak probe.

For the first step, we diagonalize the two-level system spanned by $|g\rangle$ and $|e\rangle$, setting $\alpha=\beta=\gamma_e=0$. We assume without loss of generality that the relative phase between states $|g\rangle$ and $|e\rangle$ is chosen such that $\Omega_R$ is real. This yields energy eigenvalues and eigenvectors
\begin{subequations}
\begin{eqnarray}
\label{E-pm}
E_\pm &=& \frac\hbar2 (\Delta_e + \Delta_g \pm \Omega_{\rm eff}) \\
|+\rangle &=& \cos\frac\vartheta2 |e\rangle + \sin\frac\vartheta2 |g\rangle \\
|-\rangle &=& - \sin\frac\vartheta2 |e\rangle + \cos\frac\vartheta2 |g\rangle
,
\end{eqnarray}
\end{subequations}
where the effective Rabi angular frequency $\Omega_{\rm eff}$ and the mixing angle $\vartheta$ are real-valued and must satisfy the implicit equations
\begin{subequations}
\begin{eqnarray}
\label{cos-theta}
\Omega_{\rm eff} \cos\vartheta &=& \Delta_e-\Delta_g \\
\Omega_{\rm eff} \sin\vartheta &=& \Omega_R
\label{sin-theta}
.
\end{eqnarray}
\end{subequations}
This determines a unique value of $\vartheta$ modulo $2\pi$ and it yields
\begin{eqnarray}
\label{Omega-eff}
\Omega_{\rm eff} = \sqrt{\Omega_R^2+(\Delta_e-\Delta_g)^2}
.
\end{eqnarray}

For the second step, we rewrite the mean-field model (\ref{mean-field}) in the new basis and obtain with $\beta=0$
\begin{subequations}
\begin{eqnarray}
i \frac{d}{dt} \psi_a &=& \frac{U_{\rm bg}}{\hbar} |\psi_a|^2 \psi_a 
+ 2 \psi_a^* (C_+^* \psi_+ + C_-^* \psi_- ) 
\\
i \frac{d}{dt} \psi_+ &=& C_+ \psi_a^2 
+ \left(\frac{E_+}\hbar - \frac i2 \gamma_+ \right) \psi_+
- \frac i2 \gamma_{\rm mix} \psi_- 
\\
i \frac{d}{dt} \psi_- &=& C_- \psi_a^2 
- \frac i2 \gamma_{\rm mix} \psi_+
+ \left(\frac{E_-}\hbar - \frac i2 \gamma_- \right) \psi_-
\qquad
\end{eqnarray}
\end{subequations}
with $C_+ = \alpha \sin\frac\vartheta2$, $C_- =\alpha \cos\frac\vartheta2$, $\gamma_+ = \gamma_e \cos^2\frac\vartheta2$, $\gamma_- = \gamma_e \sin^2\frac\vartheta2$, and $\gamma_{\rm mix} = -\gamma_e \sin\frac\vartheta2 \cos\frac\vartheta2$. So far we only rotated the basis and the model is still exact. We now approximate the loss as being diagonal in the states $|+\rangle$ and $|-\rangle$ by setting $\gamma_{\rm mix}=0$. Adiabatic elimination of the populations in states $|+\rangle$ and $|-\rangle$ then yields
\begin{eqnarray}
a = a_{\rm bg}+ a_+ +a_- ,
\end{eqnarray}
where the states $|+\rangle$ and $|-\rangle$ each contribute a Breit-Wigner form as a function of $E_\pm$
\begin{eqnarray}
\label{Breit-Wigner}
a_\pm = -\frac{m}{2\pi\hbar} \ \frac{|C_\pm|^2}{E_\pm/\hbar-i\gamma_\pm/2}
.
\end{eqnarray}
Thus, the states $|+\rangle$ and $|-\rangle$ each cause a single resonance and their contributions to $a$ are simply added. This corresponds to the intuitive understanding of an Autler-Townes doublet \cite{cohen-tannoudji:92}.

The above approximation $\gamma_{\rm mix}=0$ is self-consistent if the system is close to one resonance and the resonances are well separated ($\gamma_e\ll\Omega_{\rm eff}$), because in this case the states $|\pm\rangle$ have very different populations, so that a possible coherence between these populations has little effect and $\gamma_{\rm mix}$ is negligible.

\subsection{Properties of the Autler-Townes Resonances}
\label{sec-Autler-approx}
We note that $E_\pm$, $\gamma_\pm$, and $C_\pm$ in Eq.\ (\ref{Breit-Wigner}) are generally nonlinear functions of $B$. But if the resonances are narrow and well separated, then $\vartheta$ will vary only little within the width of a resonance and we can approximate $\gamma_\pm$ and $C_\pm$ as constant across a resonance. According to Eqs.\ (\ref{K2-def}) and (\ref{Breit-Wigner}), maxima of $K_2(B)$ will then occur at $E_\pm=0$. Combination with Eqs.\ (\ref{delta-g}), (\ref{delta-e}), (\ref{E-pm}), and (\ref{Omega-eff}) yields the magnetic fields $B_{\rm res}$ at which the resonances occur
\begin{eqnarray}
\label{B-res}
B_{\rm res} = B_{\rm pole} + \frac{\hbar}{2\mu_{ae}} \left( \Delta_L \pm \sqrt{ \Delta_L^2 + \frac{\mu_{ae}}{\mu_{ag}} \Omega_R^2 } \right)
.
\end{eqnarray}

The condition $E_\pm=0$ combined with Eqs.\ (\ref{K2-def}) and (\ref{Breit-Wigner}) yields the maximum of the loss rate coefficient
\begin{eqnarray}
\label{K2-tan^2}
K_2^{\rm max} = \frac{8|\alpha|^2}{\gamma_e} \left( \tan \frac\vartheta2 \right)^{\pm2} 
.
\end{eqnarray}
Insertion of $\cos^2\frac\vartheta2=\frac12(1+\cos\vartheta)$, $\sin^2\frac\vartheta2=\frac12(1-\cos\vartheta)$, Eqs.\ (\ref{E-pm}), (\ref{cos-theta}), and $E_\pm=0$ yields
\begin{eqnarray}
K_2^{\rm max} = \frac{8|\alpha|^2}{\gamma_e} \frac{\Delta_e}{\Delta_g}
.
\end{eqnarray}
Insertion of Eqs.\ (\ref{delta-g}), (\ref{delta-e}), (\ref{B-res}), and $B=B_{\rm res}$ yields
\begin{eqnarray}
\label{K2-approx}
K_2^{\rm max} = \frac{8|\alpha|^2}{\gamma_e} \ \frac{\mu_{ae}}{\mu_{ag}}
\left( 1- \frac{2\Delta_L}{\Delta_L \pm \sqrt { \Delta_L^2 + \frac{\mu_{ae}}{\mu_{ag}}\Omega_R^2 }} \right)
.
\end{eqnarray}

The width of the resonance can also be calculated. For this, we recall that the resonance is narrow and approximate $E_\pm$ as linear in $B$ around $B_{\rm res}$. As a result, Eq.\ (\ref{Breit-Wigner}) becomes a Breit-Wigner form as a function of $B$. Correspondingly $K_2(B)$ becomes a Lorentzian as in Eq.\ (\ref{K2-Lorentz}).

In this linear approximation $E_\pm=\epsilon (B-B_{\rm res})$ with $\epsilon = dE_\pm/dB|_{B_{\rm res}}$. Eqs.\ (\ref{delta-g}), (\ref{delta-e}), (\ref{E-pm}) and (\ref{Omega-eff}) yield
\begin{eqnarray}
\epsilon = \frac12\left( \mu_{ae} + \mu_{ag} \pm(\mu_{ae} - \mu_{ag}) \frac{\Delta_e-\Delta_g}{\Omega_{\rm eff}} \right)
.
\end{eqnarray}
According to Eqs.\ (\ref{K2-def}) and (\ref{Breit-Wigner}), the full width at half maximum of the Lorentzian $K_2(B)$ is $W=\hbar \gamma_\pm/\epsilon$. Using Eq.\ (\ref{cos-theta}) we obtain
\begin{eqnarray}
\frac{\hbar\gamma_e}W 
= \frac{2\epsilon}{1\pm\cos\vartheta}
= \mu_{ae} + \mu_{ag} \left( \tan \frac\vartheta2 \right)^{\pm2}
.
\end{eqnarray}
The term $\left( \tan \frac\vartheta2 \right)^{\pm2}$ can be evaluated as in Eq.\ (\ref{K2-tan^2}), yielding
\begin{eqnarray}
\label{W-approx}
W = \frac{\hbar\gamma_e}{2\mu_{ae}}
\left( 1 \pm \frac{\Delta_L} {\sqrt{ \Delta_L^2+ \frac{\mu_{ae}}{\mu_{ag}}\Omega_R^2}} \right)
.
\end{eqnarray}
We will compare these results with experimental data in Sec.\ \ref{sec-exp-Autler}.

\section{Experiment}

\begin{figure*}[t!]
\includegraphics[width=0.98\textwidth]{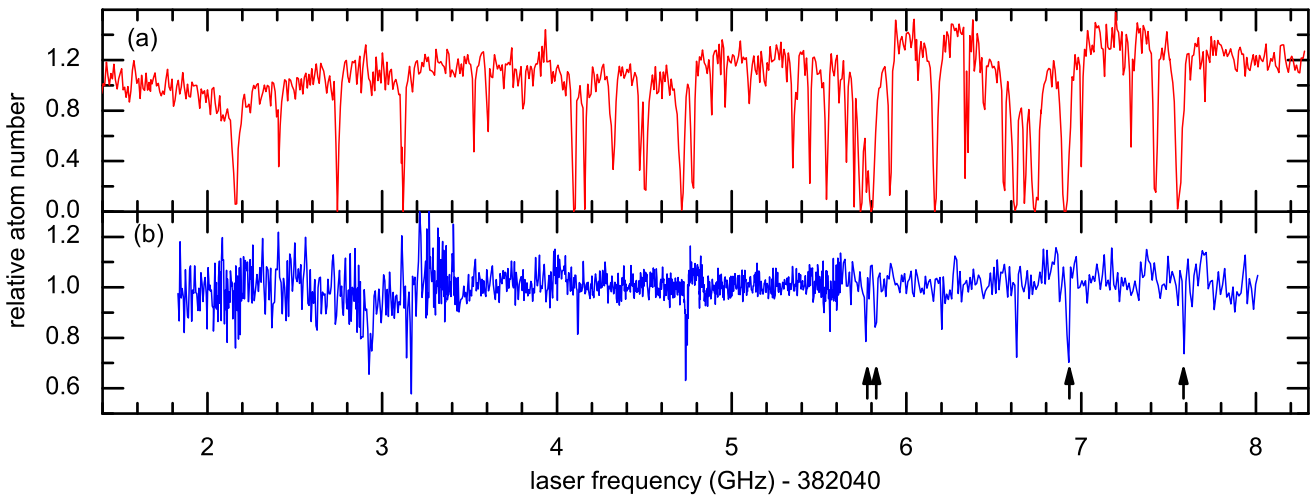}
\caption{
\label{fig-spectrum}
(Color online) Excited-state spectroscopy for $^{87}$Rb. (a) Photoassociation spectrum taken at $B=1000.0$ G. (b) Bound-to-bound spectrum taken at $B=1000.0$ G. Arrows indicate the resonances characterized in Tab.\ \ref{tab-parameters}.
}
\end{figure*}

Our experiments are performed with $^{87}$Rb using a Feshbach resonance that is characterized by the parameters \cite{marte:02,duerr:04a,duerr:04,volz:03} $B_{\rm pole}=1007.4$ G, $\Delta B=0.21$ G, $\mu_a/2\pi\hbar=1.02$ MHz/G, $\mu_{ag}/2\pi\hbar=3.8$ MHz/G, and $a_{\rm bg}=100.5 a_0$, where $a_0$ is the Bohr radius. With these parameters Eq.\ (\ref{alpha}) yields $|\alpha|/2\pi=1.8$ mHz cm$^{3/2}$. The peak density of the BEC is typically $n=|\psi_a|^2\sim 2\times 10^{14}$ cm$^{-3}$. The quantity $|\sqrt8\alpha\psi_a|=2\pi\times 0.07$ MHz can be regarded as a Rabi angular frequency. For typical parameters of our experiment, this value is small compared to $\gamma_e$ and $|\Omega_R|$, so that the adiabatic elimination performed in Sec.\ \ref{sec-a} is justified. To set the scale for $K_2$ in Fig.\ \ref{fig-doublet}, we note that $U_{\rm bg}/\hbar=4.9\times 10^{-11}$ cm$^3$/s.

\subsection{Excited-State Spectroscopy}
\label{sec-spectroscopy}

The starting point for all experiments in this paper is an essentially pure BEC of $^{87}$Rb atoms in the hyperfine state $|F,m_F\rangle=|1,1\rangle$. These atoms are held in a crossed-beam optical dipole trap with both beams operated at 1064 nm and with trap frequencies of $(\omega_x,\omega_y,\omega_z)/2\pi=(74,33,33)$ Hz. Gravity acts along the $x$ axis. The magnetic field $B$ points along the $z$ axis and is held several gauss away from the Feshbach resonance.

For the excited-state spectroscopy, we use two techniques which complement each other. The first technique is ordinary photoassociation spectroscopy. For this, we simply illuminate the BEC with photoassociation light. The light intensity is slowly increased within 80 ms to a final power of $\sim 10$ mW and held there for 100 ms. The slow increase is needed to avoid large-amplitude oscillations of the cloud shape. Next, the photoassociation light, $B$, and the dipole trap are switched off simultaneously. Finally, the remaining atom number is extracted from a time-of-flight image.

The photoassociation light is implemented as a traveling-wave laser beam with a waist ($1/e^2$ radius of intensity) of $w=0.17$ mm and a wavelength of $\sim 784.7$ nm. In order to address as many excited states as possible in the spectroscopy measurements, we let the photoassociation beam propagate along the $x$ axis and choose a specific linear polarization which corresponds to $1/3$ of the intensity in each of the polarizations $\pi$, $\sigma^+$, and $\sigma^-$. As the population is initially in the atomic state $|a\rangle$, the loss signal is typically dominated by photoassociation processes so that the technique is particularly sensitive to excited states that have a large value of the photoassociation coupling strength $\beta$. 

\begin{table}[b!]
\caption{
\label{tab-parameters} 
Parameters of four selected bound-to-bound resonances. $e$ is the elementary charge, $a_0$ the Bohr radius. $\beta$ is negligible for all these resonances.
}
\begin{ruledtabular}
\begin{tabular}{ccccc}
Polarization & $\omega_{eg}/2\pi$ & $|d_{eg}|/ea_0$ & $\gamma_e/2\pi$ & $\mu_{ae}/2\pi\hbar$ \\
 & (MHz) &  & (MHz) & (MHz/G) \\ \hline  
$\sigma^-$ & 382,045,759.4(3) & 0.24(5) & 4.4(5) & 2.2(1) \\
$\sigma^+$ & 382,045,818.2(3) & 0.29(5) & 4.7(5) & 1.7(1) \\
$\pi$      & 382,046,942.8(3) & 0.28(5) & 4.7(5) & 2.6(1) \\
$\pi$      & 382,047,581.8(3) & 0.18(5) & 5.3(5) & 2.7(1) \\
\end{tabular}
\end{ruledtabular}
\end{table}

Figure \ref{fig-spectrum}(a) shows a photoassociation spectrum at $B=1000.0$ G. We extended this photoassociation scan down to 382,037.3 GHz but did not find any further loss resonances. We determined the corresponding zero-field frequencies by performing a similar measurement at $B\sim 0$. We took data between 382,034 GHz and 382,051 GHz and found photoassociation loss resonances in the range between 382,041.8 GHz and 382,044.5 GHz. A comparison with photoassociation data from the Heinzen group shows that the excited states involved are the hyperfine and magnetic substates of the vibrational state $v=120$ in the attractive $1_g$ potential that is adiabatically connected to the $^2\! P_{3/2} + \, ^2\! S_{1/2}$ threshold \cite{tsai:pers:normal}.

Since our technique for shifting a Feshbach resonance with laser light relies on a bound-to-bound transition, not on photoassociation, we developed a second spectroscopy technique that is particularly sensitive to excited states with a large value of $\Omega_R$. The basic idea is to first use the Feshbach resonance to associate molecules into state $|g\rangle$ and then illuminate them with light that resonantly drives bound-to-bound transitions. We call this the bound-to-bound light and employ the same laser beam previously used for the photoassociation spectroscopy.

In order to avoid loss of particles due to inelastic collisions between molecules, the atoms must be loaded into a deep optical lattice before associating the molecules \cite{thalhammer:06}. The lattice has a light wavelength of 830.440 nm and a depth of $V_0\sim 20 E_r$, where $E_r$ is the atomic recoil energy. As in Ref.\ \cite{volz:06} we prepare an atomic Mott insulator, which contains exactly two atoms at each lattice site in the central region of the lattice. This core is surrounded by a shell of sites that contain exactly one atom each. After loading the lattice, the laser power of one of the dipole trapping beams is ramped to zero, as in Ref.\ \cite{volz:06}.

Next, we ramp the magnetic field slowly downward across the Feshbach resonance so that molecules are associated \cite{volz:06,koehler:06} in state $|g\rangle$ at sites that contain exactly two atoms. Sites containing three or more atoms might exist due to imperfect state preparation. Such sites are emptied by inelastic collisions as soon as molecules begin to form. Sites containing one atom are unaffected by the ramp.

Next, the bound-to-bound light is turned on for 0.2 ms at a power of $\sim 0.1 \ \mu$W. This light has the same linear polarization as for the photoassociation spectroscopy. If a molecule in state $|g\rangle$ is excited on a bound-to-bound transition, then it is likely to undergo spontaneous radiative decay into a different internal state. After turning off the bound-to-bound light, the magnetic field is ramped back across the Feshbach resonance to dissociate the molecules that remained in state $|g\rangle$. Subsequently, the optical dipole trap at 1064 nm is turned back on, the lattice depth is slowly ramped to zero, the cloud is released, $B$ is switched off, and the atom number is determined from a time-of-flight image. Molecules that were excited by the laser are not dissociated and thus not detected.

A bound-to-bound spectrum measured at $B=1000.0$ G is shown in Fig.\ \ref{fig-spectrum}(b). Comparison with the photoassociation spectrum in part (a) shows that many of the excited states are visible with both techniques. But identifying promising candidates with large $\Omega_R$ is not easily possible from part (a). The light frequency calibration for both spectra has a precision of $\sim 30$ MHz and can fluctuate within a single scan. This causes deviations in the resonance positions between the photoassociation spectrum and the bound-to-bound spectrum. In addition, the states $|a\rangle$ and $|g\rangle$ are degenerate at $B=1007.4$ G. At $B=1000.0$ G, the internal energies of a molecule in state $|g\rangle$ and a pair of atoms in state $|a\rangle$ differ by $2\pi\hbar \times 20$ MHz \cite{duerr:04}. In Fig.\ \ref{fig-spectrum} this yields a 20-MHz shift of all bound-to-bound resonances with respect to the photoassociation resonances.

Very different values of the light intensity and the illumination time were used when recording the two spectra. This tremendous difference in the sensitivity of the two methods is a result of Eq.\ (\ref{beta-small}).

In Ref.\ \cite{bauer:09} we developed a method to determine all the parameters of a bound-to-bound resonance. We now apply this method to four reasonably strong bound-to-bound resonances which are fairly close to the high-frequency end of the spectrum in Fig.\ \ref{fig-spectrum}. Results are listed in Tab.\ \ref{tab-parameters}. For these measurements, the laser producing the bound-to-bound light was beat-locked to a frequency comb, resulting in a much better precision of the frequency calibration. The polarization of each resonance was determined from a series of measurements in which the bound-to-bound light had only one of the polarizations $\pi$, $\sigma^+$, or $\sigma^-$. The latter two polarizations were implemented with the bound-to-bound beam propagating along the $z$ axis. Each resonance in Tab.\ \ref{tab-parameters} responded to only one of these polarizations.

Our choice of the light wavelength for the experiments in the following Sec.\ \ref{sec-shift} is based on the spectra obtained here. In order to achieve a large detuning from all excited states, the light must be detuned either to the left or to the right of the complete spectrum in Fig.\ \ref{fig-spectrum}. The three outermost resonances at the low-frequency end of the spectrum in Fig.\ \ref{fig-spectrum} have fairly strong photoassociation loss features, but show hardly any bound-to-bound features, which is unfortunate. The high-frequency end of the spectrum looks more promising. We therefore perform all the following experiments blue detuned from the high-frequency end of the spectrum in Fig.\ \ref{fig-spectrum}. According to Tab.\ \ref{tab-parameters}, the two strongest bound-to-bound resonances near this end of the spectrum both respond to $\pi$ polarized light. Hence, we choose $\pi$ polarization for the bound-to-bound light in all the following experiments.

We note as a side remark, that our model, unlike Ref.\ \cite{mackie:08}, assumes the existence of a direct bound-to-bound coupling term $\Omega_R$. In Ref.\ \cite{mackie:08} an indirect bound-to-bound coupling is constructed by invoking virtual transitions into the continuum of excited atom-pair states above threshold. This implies that the bound-to-bound coupling should be proportional to the photoassociation coupling $\beta$ \cite{mackie:08}. Our spectroscopy data in Fig.\ \ref{fig-spectrum} do not support this prediction of Ref.\ \cite{mackie:08}. There are strong photoassociation resonance that have hardly any bound-to-bound coupling. This indicates that the indirect coupling discussed in Ref.\ \cite{mackie:08} is negligible \cite{Autler-paper:beta}. 

\subsection{Shifting the Feshbach Resonance with Light}
\label{sec-shift}

We now use the spectroscopic information gathered above to shift the Feshbach resonance with far-detuned light. As discussed in the introduction of this paper, we minimize the effect of the dipole trap created by the bound-to-bound light by working in a deep optical lattice.

In order to measure ${\rm Re}(a)$, we first load the atoms into the lattice as described in Sec.\ \ref{sec-spectroscopy}. We then use excitation spectroscopy \cite{stoeferle:04} in the lattice, {\it i.e.}, we modulate the power of one retro-reflected lattice beam sinusoidally as a function of time around an average lattice depth of $V_0\sim 15 E_r$. The modulation amplitude is $\sim 4 E_r$. The modulation lasts for 10 or 20 ms. During the modulation, the atoms are illuminated with the bound-to-bound light and $B$ is held at a specific value close to the Feshbach resonance. The bound-to-bound light is on for a long enough time that sites containing two or more atoms are essentially emptied by light-induced inelastic collisions. The signal in the excitation spectrum that is sensitive to the modulation frequency thus stems from sites that were initially populated by one atom. For certain modulation frequencies, tunneling of an atom between two such sites is resonantly enhanced. This leads to a frequency-dependent loss of atoms and of atomic phase coherence. At the end of the modulation, we switch the bound-to-bound light off and simultaneously jump $B$ back to a value several gauss away from the Feshbach resonance. Next, the dipole trap at 1064 nm is turned back on and the lattice depth is slowly reduced to $V_0\sim 6 E_r$, where the gas is superfluid, thus restoring phase coherence between neighboring lattice sites. Finally, the dipole trap, $B$, and the lattice are simultaneously switched off. The time-of-flight image shows satellite peaks due to the restored phase coherence.

The visibility \cite{gerbier:05} of the satellite peaks displays a minimum at a modulation frequency where tunneling processes between two initially singly occupied sites are resonant. This minimum is located at a frequency $f={\rm Re}(U)/2\pi\hbar$ with the on-site interaction matrix-element $U=g\int d^3x |w({\bf x})|^4 $, where $g=4\pi\hbar^2 a/m$ and $w$ is a tight-binding Wannier function. The measurement of $f$ thus yields ${\rm Re}(a)$.

\begin{figure}[t!]
\includegraphics[width=0.9\columnwidth]{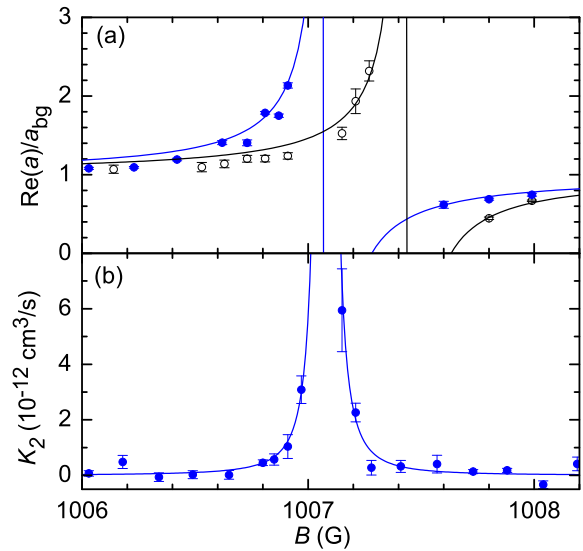}
\caption{
\label{fig-shift}
(Color online) Shifting the Feshbach resonance with laser light. (a) Elastic and (b) inelastic two-body scattering properties are shown as a function of magnetic field $B$. Experimental data in the presence (\color{blue}$\bullet$\color{black}) and absence ($\circ$) of the light are compared. The light power is 11.2 mW and the frequency is $\omega_L/2\pi=382,048,158$ MHz, which is 576 MHz blue detuned from the nearest bound-to-bound transition. The solid lines in (a) and (b) show fits of Eqs.\ (\ref{a-FR}) and (\ref{K2-Lorentz}), respectively, to the data. The Feshbach resonance is shifted by $\sim - 0.35$ G. At $B=1006.91$ G, we measure ${\rm Re}(a)/a_{\rm bg}-1\sim 1$ and $K_2\sim 1\times 10^{-12}$ cm$^3$/s.
}
\end{figure}

A sequence of such measurements for various values of $B$ yields Fig.\ \ref{fig-shift}(a). For parameters where ${\rm Re}(a)$ is reduced drastically, the system becomes superfluid and the peak in the excitation spectrum is smeared out so much that its center cannot be determined any more. Hence, this method is not applicable in this regime. For comparison, the figure also shows ${\rm Re}(a)$ measured with the same method, but in the absence of bound-to-bound light. Clearly, the position $B_{\rm res}$ of the Feshbach resonance is shifted by $\sim - 0.35$ G due to the presence of the light.

We now turn to the question, how large a loss-rate coefficient $K_2$ is associated with this shift. In order to determine $K_2$, we load the atoms into the lattice and associate molecules as described in Sec.\ \ref{sec-spectroscopy}. Right after association, we dissociate the molecules. This association-dissociation sequence serves the purpose of emptying all sites that contain three or more atoms, which will become important below. The lattice depth is $V_0\sim 20E_r$ so that tunneling is negligible. We then switch on the bound-to-bound light and simultaneously jump $B$ to a value close to the Feshbach resonance. These conditions are maintained for a variable hold time. During this hold time, the bound-to-bound light causes rapid loss of atoms in doubly occupied sites. Next, $B$ is switched to a value several gauss away from the Feshbach resonance and the bound-to-bound light is switched off. The dipole trap light at 1064 nm is turned back on and the lattice depth is slowly lowered to zero. Finally, the cloud is released and $B$ is switched off. The remaining number of atoms is extracted from a time-of-flight image.

The two-body loss during the hold time is described by the master equation of Ref.\ \cite{syassen:08}. We consider a lattice site initially occupied by exactly two atoms and we neglect tunneling between sites. The master equation then yields the density matrix $\rho= p |2\rangle\langle 2|+(1-p) |0\rangle\langle 0|$, where $p=\exp(-\Gamma t)$ is the probability that a decay at this site occurred and $|n\rangle$ denotes a Fock state with $n$ atoms. The parameter $\Gamma$ is given by \cite{syassen:08,Autler-paper:Gamma}
\begin{eqnarray}
\label{Gamma}
\Gamma=K_2 \int d^3x |w({\bf x})|^4
.
\end{eqnarray}
The decay of the total atom number $N$ in the experiment is obtained by taking the sum over a large number of isolated lattice sites, yielding
\begin{eqnarray}
\label{exponential}
N(t)=N_1+N_2 \exp(-\Gamma t)
,
\end{eqnarray}
where $N_1$ and $N_2$ are the initial atom numbers on singly and doubly occupied sites, respectively. It is crucial that there are no sites with three or more atoms, because they would give rise to an additional term that would decay more rapidly, thus making it more difficult to extract $\Gamma$ from the measured $N(t)$.

We measured $N(t)$ at a fixed hold time $t=2.1$ ms for various values of $B$ and used Eqs.\ (\ref{Gamma}) and (\ref{exponential}) to extract $K_2(B)$. Results are shown in Fig.\ \ref{fig-shift}(b). At $B=1006.91$ G, we measure ${\rm Re}(a)/a_{\rm bg}-1\sim 1$ and $K_2\sim 1\times 10^{-12}$ cm$^3$/s, which is one order of magnitude lower than our previously published result of Ref.\ \cite{bauer:09} and two orders of magnitude lower than the corresponding result reported for photoassociation resonances in $^{87}$Rb \cite{theis:04,thalhammer:05}.

We use the results of Sec.\ \ref{sec-large-detuning} to calculate the theoretical expectations from the sum of the two $\pi$ resonances in Tab.\ \ref{tab-parameters}. We thus expect $B_{\rm res}-B_{\rm pole}=-0.24$ G and $K_2=3\times 10^{-13}$ cm$^3$/s at $|B-B_{\rm res}|=\Delta B$ which in the model corresponds to ${\rm Re}(a)/a_{\rm bg}-1=\pm1$. Both experimentally observed values are somewhat larger than the expectation. This might be due to contributions from other bound-to-bound and photoassociation resonances that we did not include in this estimate.

We tried to reduce $K_2$ even further by setting the power of the bound-to-bound light to 66 mW and its frequency to 382,050,911 MHz which corresponds to a detuning of $\Delta_L/2\pi=3.33$ GHz from the nearest bound-to-bound resonance. The expected and observed shifts were $-0.35$ G and $\sim -0.65$ G, respectively. But here we observed $K_2\sim 3\times 10^{-12}$ cm$^3$/s at $|B-B_{\rm res}|=\Delta B$, which is much worse than the expectation $K_2=1\times 10^{-13}$ cm$^3$/s. As the detuning is much larger than in Fig.\ \ref{fig-shift}, other bound-to-bound and photoassociation resonances contribute even more strongly to the signal, which might explain the increased deviation between the observed values and the estimates based on the two $\pi$ resonances of Tab.\ \ref{tab-parameters}.

\subsection{Autler-Townes Doublet}
\label{sec-exp-Autler}

\begin{figure}[t!]
\includegraphics[width=0.9\columnwidth]{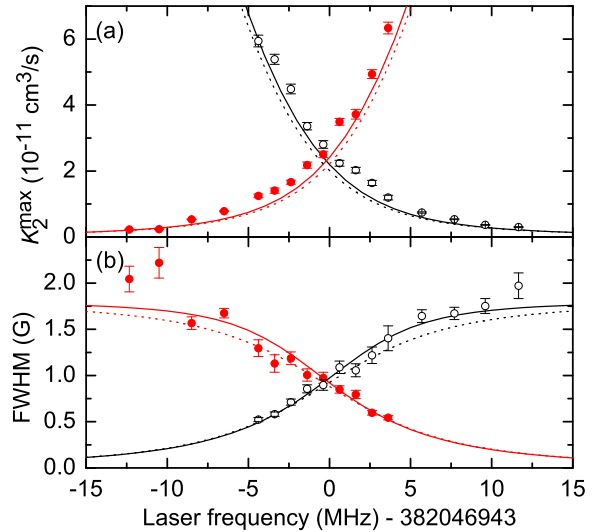}
\caption{
\label{fig-systematic}
(Color online) Systematic study of the loss resonances. $K_2(B)$ was measured for certain values of the laser frequency at a fixed laser power of 0.47 mW. (a) The maximum $K_2^{\rm max}$ and (b) the width $W$ were determined from a fit to Eq.\ (\ref{K2-Lorentz}). The experimental data for the resonances that occur at the lower ($\circ$) and higher (\color{red}$\bullet$\color{black}) value of $B$ both agree well with the predictions of the full model Eq.\ (\ref{K2}) (solid lines) which is well approximated by Eqs.\ (\ref{K2-approx}) and (\ref{W-approx}) (dotted lines).
}
\end{figure}

Finally, we compare the theoretical results for the width and height of the Autler-Townes resonances in $K_2(B)$ from Sec.\ \ref{sec-Autler-approx} with experimental results. We measured Autler-Townes doublets in $K_2(B)$, as shown in Fig.\ \ref{fig-doublet}(b). We fit a Lorentzian (\ref{K2-Lorentz}) to each of the two peaks in the experimental data. The best-fit values for $K_2^{\rm max}$ and $W$ are shown in Fig.\ \ref{fig-systematic}. The experimental procedure and parameters are identical to Fig.\ 4 of Ref.\ \cite{bauer:09}; see this reference for details.

The experiment is performed in a regime where $|\Omega_R| \gg \gamma_e$ so that the Autler-Townes model of Sec.\ \ref{sec-Autler-approx} is expected to be a good approximation. The dotted lines show the corresponding predictions (\ref{K2-approx}) and (\ref{W-approx}). They agree well with the experimental data.

For comparison, we numerically determined the peaks in $K_2(B)$ from the full model (\ref{K2}) with the parameters of Tab.\ \ref{tab-parameters}. The corresponding maximum values $K_2^{\rm max}$ are shown as solid lines in Fig.\ \ref{fig-systematic}(a). As the full model (\ref{K2}) does not predict Lorentzian lines, a direct comparison with the width $W$ is not straightforward. We decide to use the second derivative of $K_2(B)$ at the maximum for a comparison. The solid lines in Fig.\ \ref{fig-systematic}(b) therefore show the values of
\begin{eqnarray}
W = \left(- \frac{1}{8K_2} \ \frac{d^2 K_2}{dB^2} \right)^{-1/2}
\end{eqnarray}
at the peaks calculated from the full model (\ref{K2}). If the peaks in the model were Lorentzian, this would yield the width $W$. The solid lines also agree well with the experimental data and with the dotted lines.

\section{Conclusion}

To summarize, we improved our recently developed scheme for shifting a magnetic Feshbach resonance with laser light by exploring the regime of even larger detuning and laser power. We demonstrated that the light-induced loss rate can be reduced by one order of magnitude compared to our pervious work \cite{bauer:09}. The measurements required excited-state spectroscopy and an optical lattice. We also presented a detailed discussion of a model that describes our experimental data.

\acknowledgments
We thank B. Bernhardt and K. Predehl for providing light from their frequency comb. We acknowledge fruitful discussions with T.\ Bergeman, D.\ Heinzen, and C.-C.\ Tsai. This work was supported by the German Excellence Initiative through the Nanosystems Initiative Munich and by the Deutsche Forschungsgemeinschaft through SFB 631.

%\bibliographystyle{unsrtSDAPS}
%\bibliography{C:/BibTeX/BEC}

\begin{thebibliography}{10}
\bibitem{Autler-paper:beta}
There is a correlation in the observed line strengths, insofar as almost every
  resonance with a strong bound-to-bound feature correlates with a fairly
  strong feature in the photoassociation spectrum. But this does not imply that
  these lines have a large $\beta$. Instead, Eq.\ (\ref{K2}) shows that
  $\Omega_R$ alone (even for $\beta=0$) can create a strong loss feature for
  atoms in state $|a\rangle$ illuminated with light resonant with the
  photoassociation transition, as long as $\Delta_g$ is not too large.
\bibitem{Autler-paper:Gamma}
Here we use $a$ and $K_2$ from Eqs.\ (\ref{a}) and (\ref{K2}). We derived these
  equations assuming that the sample is Bose condensed which is not the case in
  a deep optical lattice. But $a$ and $K_2$ express only properties of two-body
  scattering processes. They are unaffected by the many-body state of the
  system.
\end{thebibliography}

\end{document}